\documentclass[sigconf,nonacm]{acmart}

\usepackage[english]{babel}
\usepackage{blindtext}
\usepackage{booktabs}
\usepackage{graphicx}

\acmConference{}{}{}
\renewcommand\footnotetextcopyrightpermission[1]{} % removes footnote with conference information in first column
\setcopyright{none}

\usepackage{subfigure}
\usepackage{enumitem}
\usepackage{color, colortbl}
\usepackage{url}
\usepackage[font=small]{caption}
\usepackage{algorithm}
\usepackage{algorithmicx}
\usepackage{algpseudocode}

\definecolor{DarkGray}{gray}{0.95}
\definecolor{LightGray}{gray}{0.9}

\newcommand{\myitem}[1]{\vspace{0.25\baselineskip}\noindent\textbf{#1}}%\vspace*{0.04in}
\begin{document}

\title{Benchmarking Graph Neural Networks for Internet Routing Data}
% \title{Benchmarking GNNs for Internet routing data}

\author{Dimitrios P. Giakatos}
\email{dgiakatos@csd.auth.gr}
\affiliation{%
  \institution{Aristotle University of Thessaloniki}
  \country{Greece}
}

\author{Sofia Kostoglou}
\email{sofikost@csd.auth.gr}
\affiliation{%
  \institution{Aristotle University of Thessaloniki}
  \country{Greece}
}
\author{Pavlos Sermpezis}
\email{sermpezis@csd.auth.gr}
\affiliation{%
  \institution{Aristotle University of Thessaloniki}
  \country{Greece}
}
\author{Athena Vakali}
\email{avakali@csd.auth.gr}
\affiliation{%
  \institution{Aristotle University of Thessaloniki}
  \country{Greece}
}

% \author{Dimitrios P. Giakatos, ~Sofia Kostoglou, ~Pavlos Sermpezis, ~Athena Vakali}
% \email{{dgiakatos, sofikost, sermpezis, avakali}@csd.auth.gr}
% \affiliation{%
%   \institution{Aristotle University of Thessaloniki}
%   \country{Greece}
% }

\renewcommand{\shortauthors}{D.P. Giakatos, S. Kostoglou, et al.}

\begin{abstract}
The Internet is composed of networks, called Autonomous Systems (or, ASes), interconnected to each other, thus forming a large graph. While both the AS-graph is known and there is a multitude of data available for the ASes (i.e., node attributes), the research on applying graph machine learning (ML) methods on Internet data has not attracted a lot of attention. In this work, we provide a benchmarking framework aiming to facilitate research on Internet data using graph-ML and graph neural network (GNN) methods. Specifically, we compile a dataset with heterogeneous node/AS attributes by collecting data from multiple online sources, and preprocessing them so that they can be easily used as input in GNN architectures. Then, we create a framework/pipeline for applying GNNs on the compiled data. For a set of tasks, we perform a benchmarking of different GNN models (as well as, non-GNN ML models) to test their efficiency; our results can serve as a common baseline for future research and provide initial insights for the application of GNNs on Internet data.
\end{abstract}

\maketitle

\section{Introduction}

The Internet is a network of networks, which are called Autonomous Systems (or, ASes). Today there exist  more than 100k ASes originating IP prefixes in the Internet routing table, which are connected to each other through private or public peering links. Representing the ASes as nodes and their interconnections as edges, results in a large and sparse (density $<$ 0.01\%) graph.

Since ASes and their interconnections play a significant role for network operations, Internet policies, routing optimization, etc., there has been many efforts to characterize these networks. Hence, there exist rich datasets with information about ASes (open datasets~\cite{CAIDA2022ASRANK,CAIDA2022ASREL,IHR,ASDB}, self-declared databases~\cite{peeringdb}, data from custom measurements, etc.).

These datasets with AS attributes have been used by several works employing (traditional) ML methodologies for various applications~\cite{ding2016BGPanomalies,dai2019-ml-bgp-anomalies,hoarau2021graph-representation-bgp,sanchez2019comparing-ml-bgp,Cho2019BGPHC,testart2019profiling}. One would expect that with the advent of Graph Neural Networks (GNNs) many works would exploit the known AS-graph structure along the AS attributes to devise GNN-based methodologies for problems related to Internet routing and operations. However, there only exist a few efforts generating graph embeddings~\cite{shapira2022bgp2vec, shapira2020hijack-asn-embedding}, and, in fact, they are not based on GNNs (but on methods from the natural language processing field) and they do not take into account the node attributes (but only the graph structure).

While there can be many reasons behind this lack of GNN-based works for Internet routing data (and it is out of our scope to investigate them), a main challenge for applying GNNs on Internet data is that significant expertise is needed in both domains: namely, a researcher needs (i) rich Internet data and (ii) a good understanding of advanced deep learning techniques and graph theory concepts. On one hand, it may be straightforward for Internet researchers to access sources of Internet data (which are typically well known within this community), but it may be a more tedious task for researchers of other domains (e.g., more focused to GNNs) to compile a rich dataset that would be needed by a GNN architecture. On the other hand, while there are widely used and well documented libraries (pytorch geometric~\cite{pyg}, dgl~\cite{dgl}, etc.) that have made access to GNNs easy, there are many intricacies in the application of GNNs to Internet data (e.g., imbalanced data, heavy tailed distributions, etc.), which render their efficient application a non-trivial task for an Internet-focused researcher.

Motivated by the aforementioned observation, in this paper we aim to facilitate research with GNNs on Internet data through the following contributions:
\begin{itemize}[leftmargin=*,nosep]
    \item \textbf{Dataset:} We compile a \textit{rich dataset} of Internet data that can be used as input to GNN models (Section~\ref{sec:dataset}). Specifically, we collect from multiple online sources a set of 19 AS attributes, including both numerical and categorical variables. We then preprocess the data and transform them to a format that is readily available to be used as input to GNNs (e.g., all values normalized in [0,1]). 
    
    The compiled dataset not only offers easy access to researchers, but it also serves as a \textit{benchmark dataset}. The lack of benchmark datasets, has been identified as a key barrier that challenge ML research in networking applications~\cite{casas2020two}. Having a common dataset, on which different ML approaches are applied and compared (e.g., similarly to the ImageNet~\cite{deng2009imagenet} and CIFAR-10~\cite{krizhevsky2009learning} datasets in computer vision), can further boost GNN research on Internet data.

    \item \textbf{GNN benchmarking \& initial insights:} We test several GNN, graph-ML, and (non-graph) ML models on the compiled dataset, for several downstream tasks (Section~\ref{sec:methodology}). Our goal is not to propose a specific GNN architecture, and thus we refrain from extensive model optimization. Hence, we use a basic architecture and hyperparameter tuning for all models, and we produce initial results which can serve as a point of reference (e.g., baselines) for future research. Our experimental results (Section~\ref{sec:results}) provide initial insights about the efficiency of GNNs on Internet data related tasks (e.g., the role of graph structure and/or node attributes for different tasks), and reveal several challenges.
    \item \textbf{Open data and code:} We make publicly available the compiled dataset and our code (using a popular GNN library~\cite{wang2019deep,dgl})\footnote{As well as, all the experimental results of the paper, for reproducibility purposes.} in~\cite{gnn-internet-data-repository}.
\end{itemize}

\section{Dataset}\label{sec:dataset}

\begin{figure}
    \centering
    \includegraphics[width=\linewidth]{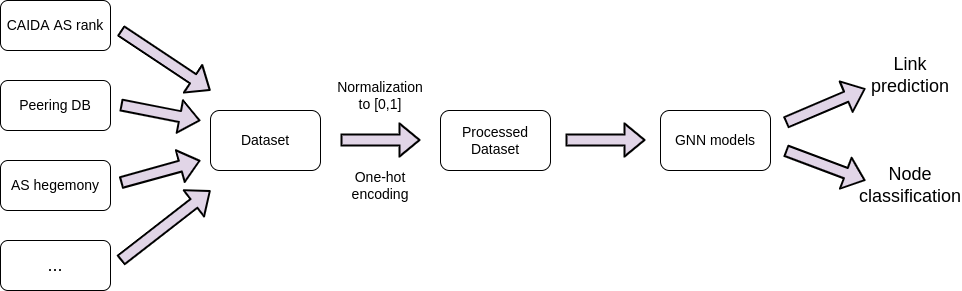}
    \caption{Overall methodology pipeline.}
    \label{fig:methodology}
\end{figure}

In this section, we present the data sources (Section~\ref{sec:data-sources}) and the preprocessing (Section~\ref{sec:data-processing}) we applied on the data to generate the compiled dataset. The overall methodology is depicted in Fig.~\ref{fig:methodology}.

\subsection{Data sources}\label{sec:data-sources}

Each network or Autonomous System (AS) can be characterized by a multitude of features, such as, location, connectivity, traffic levels, etc.. We collect data from multiple online (public) data sources to compile a dataset, which contains multiple information for each AS.

The first three data sources are widely used by Internet researchers and operators for multiple purposes:
\begin{itemize}[leftmargin=*]
    \item CAIDA AS-rank~\cite{CAIDA2022ASRANK}: various information about ASes, such as, location, network size, topology, etc.
    \item CAIDA AS-relationship~\cite{CAIDA2022ASREL}: a list of AS links (i.e., edges), which are used to build the AS-graph.
    \item PeeringDB~\cite{CAIDA2022PDB, peeringdb}: online database, where network operators register information about the connectivity, network types, traffic, etc., of their networks
    % \item RIPE Atlas probes~\cite{RIPEAtlas}
    % \item RIPE RIS route collectors~\cite{RIPERISRCs}
    % \item RouteViews route collectors~\cite{RouteViewsRCs}
\end{itemize}
We also use the following sources that provide data related to the routing properties of ASes and their business types:
\begin{itemize}[leftmargin=*]
    \item AS hegemony~\cite{AShegemony}
    \item Country-level Transit Influence (CTI)~\cite{Country-levelTransitInfluence}
    \item ASDB~\cite{ASDB}
\end{itemize}

From the above sources, we collect the most relevant attributes per AS, resulting to a dataset of 19 attributes/features (see Table~\ref{tab:features} for the detailed list). For ease of analysis, in the online repository~\cite{gnn-internet-data-repository} we also provide a visual exploratory data analysis with the detailed distributions of all attributes. 

\begin{table*}[]
    \centering
    \caption{Summary of AS attibutes/features in the compiled dataset.}
    \label{tab:features}
    \begin{small}
    \begin{tabular}{l|l|c|c}
         \textbf{Feature} &  \textbf{Description} & \textbf{Data type} & \textbf{Source}\\
         \hline
         \rowcolor{DarkGray}
         RIR region & Regional Internet registry & Categorical (6 categories)& \cite{CAIDA2022ASRANK} \\
         \rowcolor{LightGray}
         Customer cone (ASNs) & Number of ASNs in the \textit{customer cone} &  Numerical $\in [1, 48790]$ & \cite{CAIDA2022ASRANK} \\
         \rowcolor{DarkGray}
         Customer cone (prefixes) & Number of IP prefixes in the \textit{customer cone} & Numerical $\in [0,737792]$ & \cite{CAIDA2022ASRANK} \\
         \rowcolor{LightGray}
         Customer cone (addresses) & Number of IP addresses in the \textit{customer cone} & Numerical $\in [0,2090939967]$ & \cite{CAIDA2022ASRANK} \\
         \rowcolor{DarkGray}
         \#Neighbors & Total number of neighbors (in \# of ASNs) & Numerical $\in [0,9547]$ & \cite{CAIDA2022ASRANK} \\
         \rowcolor{LightGray}
         \#Customers & Total number of customers (in \# of ASNs) & Numerical $\in [0,6505]$ & \cite{CAIDA2022ASRANK} \\
         \rowcolor{DarkGray}
         \#Peers & Total number of peers (in \# of ASNs) & Numerical $\in [0,7516]$ & \cite{CAIDA2022ASRANK} \\
         \rowcolor{LightGray}
         \#Providers & Total number of providers (in \# of ASNs) & Numerical $\in [0,133]$ & \cite{CAIDA2022ASRANK} \\
         \rowcolor{DarkGray}
         Location-continent & Registered location of the headquarters of the ASN  & Categorical (6 categories)  & \cite{CAIDA2022ASRANK} \\
         \rowcolor{LightGray}
         Traffic ratio (PDB) & Type of traffic ratio (e.g., inbound, outbound, balanced) & Categorical (6 categories) & \cite{CAIDA2022PDB, peeringdb} \\
         \rowcolor{DarkGray}
         Scope (PDB) & Regional scope of the AS (e.g., regional, global, Europe) & Categorical (10 categories)  & \cite{CAIDA2022PDB, peeringdb} \\
         \rowcolor{LightGray}
         Network type (PDB) & Network type (e.g., ISP, content provider, enterprise) & Categorical (11 categories) & \cite{CAIDA2022PDB, peeringdb} \\
         \rowcolor{DarkGray}
         Peering policy (PDB) & Peering policy (e.g., open, selective, restrictive) & Categorical (4 categories)  & \cite{CAIDA2022PDB, peeringdb} \\
         \rowcolor{LightGray}
         \#IXPs (PDB) & Number of IXPs where the AS is present & Numerical $\in [0,288]$ & \cite{CAIDA2022PDB, peeringdb} \\
         \rowcolor{DarkGray}
         \#facilities (PDB) & Number of interconnection facilities where the AS is present & Numerical $\in [0,768]$ & \cite{CAIDA2022PDB, peeringdb} \\
         \rowcolor{LightGray}
         AS hegemony & Metric measuring avg. fraction of routing paths crossing an AS& Numerical $\in [0,0.2]$ & \cite{IHR}\\
         \rowcolor{DarkGray}
         CTI top & "Country-level Transit Influence" metric of an AS  & Numerical $\in [0,0.95]$ & \cite{Country-levelTransitInfluence} \\
         \rowcolor{LightGray}
         CTI origin & Percentage of addresses initiated by an AS in a country & Numerical $\in [0,97.39]$  & \cite{Country-levelTransitInfluence} \\
         \rowcolor{DarkGray}
         ASDB & Industry type of the organization of the AS & Categorical (17 categories)  & \cite{ASDB} 
    \end{tabular}
    \end{small}
\end{table*}

% \pavlos{@Sofia: add figure depicting the methodology}

\subsection{Data preprocessing}\label{sec:data-processing}

The collected data are highly heterogeneous, including both numerical and categorical attributes. Moreover, numerical attributes take values in different ranges, and some of them span ranges several orders of magnitude larger than others (see Table~\ref{tab:features}). Since, it is well known that non-homogeneous data values can impact the performance of deep learning models, we need to preprocess the data. In the following we describe the transformation we apply to each type of attributes to generate a dataset with normalized attributes taking values in the interval [0,1].

\myitem{Categorical features.}
% \pavlos{@Sofia: write very briefly what we do (one hot encoding). Give an example, e.g., for RIR, we define a vector of size X (each element corresponds to each region \{ARIN, RIPE, ...\}. The vector of each AS has one element with value 1, and all other values are 0, e.g., ASxxx is in region XXX, so the vector is [0,0,1,0,.....]}
For every categorical feature, one-hot encoding is applied. In the one-hot encoding technique, a new feature is created for every value of the categorical feature. For example, the "Location-continent" feature contains 6 values (Africa, Asia, Europe, N. America, S. America, Oceania), which means that after the one-hot enconding 6 new numerical columns are created; hence, an AS located in Europe will have a value of 1 in the respective new feature for Europe, and a value of 0 in the other 5 new features that correspond to the other continents.

%Each categorical features is mapped to 0 or 1, which correspond to presence or absence of that category, respectively.
% This means that new columns are created in the dataframe, each one for every possible value of the categorical feature.
% For example, the AS rank continent feature contains 6 continents, including North America, Europe, Asia, South America, Oceania, Africa and one more value, which corresponds to null (no continent recorded).

\myitem{Numerical features.} As it can be seen in Table~\ref{tab:features}, some numerical attributes take values in very large ranges (e.g., the customer cone of ASes spans from 1 to more than 48k ASNs). Also, for many of these attributes the values for different ASes are not distributed uniformly, but they have a heavy tail distribution (e.g., almost 95\% of ASes have a customer cone of 1 ASN). To alleviate this large heterogeneity and variability of the numerical features, we perform the following transformations.
\begin{itemize}[leftmargin=*]
    \item First, for every numerical feature, except for the \textit{AS hegemony} and the \textit{CTI top} features that only take values less than 1, we apply a logarithmic transformation to decrease their variability, as follows: $x\rightarrow\log(x+1)$. 
    
    \item Then, we normalize all numerical feature according to the Min-Max scaling method: $x\rightarrow \frac{x-min(x)}{max(x)-min(x)}$. As a result, all the resulting values are in the range of $[0,1]$. 
\end{itemize}

% First, for every numerical feature, except for the \textit{AS hegemony} and the \textit{CTI top} features that only take values less than 1, we apply a logarithmic transformation to decrease their variability, as follows: $x\rightarrow\log(x+1)$. Then, we normalize all numerical feature according to the Min-Max scaling method: $x\rightarrow \frac{x-min(x)}{max(x)-min(x)}$. As a result, all the resulting values are in the range of $[0,1]$.

% Below, the formula of the Min-Max feature scaling method is given:
% \begin{equation}
% x\;'=\frac{x-min(x)}{max(x)-min(x)}
% \end{equation}
% where $x$ is an original value and $x\;'$ is the normalized value.

% For example, the AS rank provider values are in range of [0, 133] at first. For the rescaling, 0 is substracted by every value of the feature, and the result is diveded by 133.

\myitem{Graph preprocessing.} The AS graph contains a large number of leaf nodes (i.e., edge networks with a single upstream). These nodes are of limited interest in the ML downstream tasks we consider (see Section~\ref{sec:tasks}), namely, for (i) link prediction: they only have a single link, and (ii) node classification: the characteristics/classes we consider can be easily inferred for edge networks. Moreover, taking them into account would lead to a graph structure that is more challenging to be captured by a GNN or graph-ML model. Hence, we preprocess the graph and remove all nodes with degree equal to one (and repeat two more times this process); the resulting graph has around 46K nodes and 434K edges.

\section{GNN Benchmarking Methodology}\label{sec:methodology}

To benchmark GNNs on the compiled dataset, we use a set of GNN, graph-ML, and traditional ML models (Section~\ref{sec:models}), and design the downstream tasks on which the efficiency of the models will be tested (Section~\ref{sec:tasks}).

\subsection{Models}\label{sec:models}
\myitem{GNN models:} We consider three widely used GNN models. 

\textit{\textbf{GraphSAGE}}~\cite{hamilton2017inductive} learns a function (neural network) that generates embeddings for a node by sampling and aggregating node features from its local neighborhood. The embeddings capture both the local graph structure of a node and the feature distribution of its neighborhood.

\textit{\textbf{GCN}}~\cite{kipf2016semi} (Graph Convolutional Networks) is the graph analogy to CNNs for images. It uses a spectral convolution of the graph, which leverages the Laplacian matrix in a trainable function that aggregates features from neighboring nodes. 

\textit{\textbf{GAT}}~\cite{velivckovic2017graph} (Graph Attention Networks), similarly to the above models, aggregates node features from a node's neighborhood. However, it can learn different weights for different nodes in a neighborhood, thus capturing the different levels of importance that the neighbors of a node may have.

For each model, we build a basic architecture that comprises two GNN layers (with a 32-dimensional output), followed by an MLP layer. We refrain for extensive tuning, and consider fully connected layers (no dropout), a learning rate of 0.01 and a few hundreds epochs per model.

\myitem{Graph embedding models (non-GNNs):} The goal of graph embeddings is to map the nodes of a graph to an embedding space (i.e., a vector) of lower dimensions. %This is based on the idea that nodes similar in the graph will yield embeddings.  The embedding space is just a vector corresponding to every node of the graph. 
In our experiments, we use two methods that generate node embeddings based on the graph structure, but \textit{without taking into account the node attributes}.  

\textbf{\textit{Node2vec}}~\cite{grover2016node2vec} is based on the popular word2vec~\cite{mikolov2013distributed} method used in Natural Language Processing to represent words in a text with vectors. Node2vec generates a collection of random walks on the graph, starting each time from different nodes. Those random walks are lists of nodes, which act similarly to sentences in a text (where the nodes are the words). %, having started from a certain node, and then moving to another node based on a probabilistic outcome. Node2vec starts, having a graph as input. From this graph, a collection of random walks are generated, that can be compared to sequence of words, having every node as a word. Node2vec extends the concept of random walks, by using biasing parameters of DFS or BFS algorithms parameters.  Those biased random walks are given as input to a skip-gram model, which the Word2vec algorithm uses on sentences and words. The skip-gram is a shallow neural network consisting of one hidden layer, and is trained to predict the probabilty of a certain word to be present, when an input word is there. Training set contains pairwise combinations of the target word with all the other words in a defined window that slides through every sentence. So, in the same way, the skip-gram model uses the random walks to yield an embedding for every node. 
Those lists are used as input to train a skip-gram model (shallow neural network)
to predict the probability of a certain word/node to be present in a sentence when an input word/node is present.

\textbf{\textit{Bgp2vec}}~\cite{shapira2022bgp2vec} is a method designed specifically to generate AS embeddings. It is also based on word2vec, but rather than performing random walks, it uses BGP announcements collected by route collectors~\cite{RouteViewsRCs}, and in particular the AS-paths in them as the lists of node sequences.  The model is trained over a large corpus of AS paths, and learns to characterize an AS by its context, i.e., its neighboring ASes. 

In both models we generate embeddings of size 16. In the node2vec model we use 20 random walks of a walk length equal to 4, which is around the average AS path length in the Internet.\footnote{The other detailed parameters for the models are
\begin{itemize}
    \item \textit{node2vec}: $p=1, q=1, workers=4, window\_size=5, epochs=1, lr = 0.05, min\_count=1$.
    \item \textit{bgp2vec}: $negative=5, epochs=3, window=2, shuffle=False$ 
\end{itemize}
}

% In the first stage a d-dimensional continuous vector representation for each ASN is extracted.
% As in word embeddings of Word2vec, the network is trained over a large corpus of AS paths. A similar Skip-Gram model is deployed, such that for each ASN in a
% certain AS-path ASNs are predicted within a certain range before and after the current ASN. As a result, the model learns to characterize an ASN by its context,
% meaning its neighboring ASNs.
% The intention is to learn $d$-dimensional embeddings of ASNs from paths $P$ in the dataset. First, a vocabulary is built of all the ASNs in the dataset,
% and randomly initializing the embeddings for all ASNs in the vocabulary.

% \pavlos{@Sofia: write one short paragraph about node2vec, and one short about bgp2vec}

\myitem{Traditional ML model:} Finally, we use a Random Forest (RF) model with 100 trees as a baseline, which takes as input the node features but \textit{neglects the graph structure}. 

\subsection{Learning tasks}\label{sec:tasks}

We selected two (sets of) tasks for the benchmarking:

\textbf{\textit{Link prediction}} between the nodes of the graph, i.e., inference of AS-AS links. It is well-known that our view of the AS-graph is incomplete, and there is evidence that many links (in particular, peer-to-peer links at Internet eXchange Points, or IXPs) cannot be seen by the public measurements from which we construct the known AS-graph. Therefore, being able to predict links could be important for several Internet routing use cases.

\textbf{\textit{Node classification}}: We predict attributes of ASes that are declared in the PeeringDB database~\cite{peeringdb}. Network operators voluntarily register the information of their ASes in PeeringDB, which leads to incomplete knowledge; in our dataset, only around 25\% of the ASes have information about the PeeringDB attributes. Nevertheless, knowing the PeeringDB attributes can be helpful for a number of operational/policy/economic reasons, and has been identified as a need by the network operators community.

Link prediction and node classification tasks have the same pipeline: We get the node embeddings from GNN models (GraphSAGE, GCN, GAT) or the node2vec/bgp2vec models, and feed them to an MLP network to get the predictions\footnote{In the case of the random forest, we directly get the predictions.}. We use a binary cross entropy as a loss function for the link prediction task and the cross entropy loss (due to the many classes) for the node classification tasks\footnote{Since labels are not known for all nodes, for the classification task, while we consider all nodes in the generation of the embeddings (GNN), we train/test the MLP by using only the values for the nodes with labels (i.e., by masking the nodes without labels).}.

Finally, since (i) the AS-graph is very sparse and (ii)
for some categorical features there are classes with very few samples, we balance the train datasets by selecting equal number of existing/non-existing links and class samples. In the case of classes with very few samples (less than 500), we group these classes together, in order to get enough samples to train the GNN models. The final classes for each categorical features are given in Table~\ref{table:classes}.

\begin{table}[h]
    \centering
    \caption{Classes and number of samples in each class for the categorical features used in the node classification tasks.}
    \label{table:classes}
    \begin{small}
    \begin{tabular}{l|l|r}
     \textbf{Feature}    & \textbf{Classes} & \textbf{nb. of samples}  \\
     \hline
     Traffic ratio (PDB)& Balanced & 1546 \\
     {}& Heavy Inbound & 319 \\
     {}& Heavy Outbound & 202 \\
     {}& Mostly Inbound & 1403 \\
     {}& Mostly Outbound & 517 \\
     {}& Not Disclosed & 1769 \\\hline
     Scope (PDB)& Asia Pacific & 530 \\
     {}& Europe & 869 \\
     {}& Global & 600 \\
     {}& North America & 388 \\
     {}& Not Disclosed & 1526 \\
     {}& Regional & 1515 \\
     {}& Other & 380 \\\hline
     Network type (PDB)& Cable/DSL/ISP & 1910 \\
     {}& Content & 649 \\
     {}& Enterprise & 348 \\
     {}& NSP & 998 \\
     {}& Not Disclosed & 1315 \\
     {}& Other & 579 \\\hline
     Peering policy (PDB)& Open & 4379 \\
     {}& Selective & 1018 \\
     {}& Other & 231 \\
    \end{tabular}
    \end{small}
\end{table}

% Link prediction and node classification tasks have the same pipeline. The pipeline contains three strategies:
% 1.	GNN + MLP: We use three GNN models (GraphSAGE, GCN, GAT) to get the node embeddings and then we use them as input to MLP model to get the predictions.
% 2.	Node embeddings + MLP: We use the node embeddings as inputs to MLP model to get the predictions.
% 3.	Random Forest: We use the node features in Random Forest to get the predictions.
% For 1 and 2, we use binary cross entropy as a loss function for the link prediction task and the cross entropy loss for the node classification task.
% For 1 to 3 we use the same metrics per task. In link prediction task we use the Area Under the Receiver Operating Characteristic Curve, precision and recall metric. In node classification task we use Accuracy, macro f1 score, mean squared error and mean absolute error metric.

\section{Results}\label{sec:results}

\subsection{Link prediction}
Table~\ref{tab:results-link-prediction} presents the link prediction results, reported as the average AUC score (over 10 different models, with the same hyperparameters, and random weight initialization). The largest the AUC score (in the interval [0,1]) the more accurate the prediction; a score of 0.5 corresponds to a (dummy) random predictor. Since, the underlying graph is very sparse, we also present the Recall ($\frac{\#TP}{\#P}$) and Precision ($\frac{\#TP}{\#TP+\#FP}$) metrics that focus on the "true positive" (TP) samples, i.e., existing links that are predicted correctly, and their fractions with respect to the number of all existing links ($\#P$) and all links predicted correctly or incorrectly ($\#TP+\#FP$).

We can see that the GraphSAGE and GCN models are very efficient in predicting links between nodes, whereas the GAT model has poor capacity. While the highest AUC score is achieved by the random forest (RF), its Recall value is very low: this indicates that while RF does not mispredict non-existing links (low false positives, FP) it is only able to predict 1/4 of the actual links. The graph-ML models (and, in particular, bgp2vec~\cite{shapira2022bgp2vec}) achieve also a high performance. On one hand, this shows that only the structure of the AS-graph (without node features) can help us to predict and characterize links, as already shown in~\cite{shapira2022bgp2vec}. However, the fact that the GCN model with very light tuning can outperform bgp2vec, indicates that taking into account node features can be promising for tasks related to link prediction and characterization (e.g., inference of peering relationships). 

\begin{table}[h]
    \centering
    \caption{Results for the link prediction task: average AUC, Recall, and Precision metrics over 10 runs per model.}
    \label{tab:results-link-prediction}
    \begin{tabular}{c|ccc}
         \textbf{Model} & \textbf{AUC} &\textbf{Recall ($\frac{\#TP}{\#P}$)} & \textbf{Precision ($\frac{\#TP}{\#TP+\#FP}$)} \\
         \hline
         GraphSAGE & 94.7\% & 82.7\% & 86.8\% \\
         GCN & 95.3\% & 85.5\% & 95.9\% \\
         GAT & 64.4\% & 24.2\% & 28.4\% \\
         % GIN & 50.0\% & \% & \% \\
         \hline
         node2vec & 86.5\% & 82.7\% & 95.5\% \\
         bgp2vec &  93.0\% & 85.5\% & 91.5\% \\
         \hline
         Rnd. forest & 96.2\% & 24.2\%  & 96.3\%\\
    \end{tabular}
\end{table}

In Table~\ref{tab:links-detailed} we do a deeper inspection, to understand what types of links are easier to predict. We consider the GraphSAGE model and group nodes in three categories based on the size of their neighborhood: nodes with \textit{low} ($<10$), \textit{medium} ($\in[10, 20)$), and \textit{large} ($\geq 20$) number of neighbors; around 80\% of nodes belong to the first group and around 10\% to each of the other groups. We can see that predicting links between nodes of high degrees is easy, whereas links between nodes with "low" number of neighbors (i.e., the majority of nodes) are more difficult to be inferred. This indicates a challenge and a need for efficient designs of GNN models that focus on the sparse parts of the AS-graph. 

\begin{table}[h]
    \centering
    \caption{Detailed link prediction results (Recall / Precision) for node pairs of different size of neighborhoods.}
    \label{tab:links-detailed}
    % \begin{small}
     \begin{tabular}{c|ccc}
         {}     & Low & Medium & High  \\
         \hline
         Low    &31.6\% / 89.8\% & 80.7\% / 98.2\% & 67.2\% / 90.5\%\\
         Medium &  & 94.4\% / 98.1\% & 94.4\% / 95.4\% \\
         High   &  &  & 99.5\% / 99.0\% \\
    \end{tabular}
    % \begin{tabular}{c|ccc}
    %      {}     & Low & Medium & High  \\
    %      \hline
    %      Low    & 65.7\% / 31.6\% / 89.8\% & 90.2\% / 80.7\% / 98.2\% & 82.4\% / 67.2\% / 90.5\%\\
    %      Medium &  & 94.8\% / 94.4\% / 98.1\% & 85.8\% / 94.4\% / 95.4\% \\
    %      High   &  &  & 66.9\% / 99.5\% / 99.0\% \\
    % \end{tabular}
    % \end{small}
\end{table}

\subsection{Node classification}

Table~\ref{tab:results-node-classification} presents the results of the node classification, where we try to infer different attributes of ASes related to the PeeringDB (see column names). We present the average Accuracy (ACC) and the F1 score metrics. We stress that for each attribute there are several categories (cf. Table~\ref{tab:features}), i.e., the problems in hand are \textit{multi-class} classification problems.

We can see that the best performing model differs among attributes. GraphSAGE performs consistently well, while GAT has now a comparable performance to other GNN models. Nevertheless, we believe that there may be significant room for improvement in future work, e.g., through optimization of GNN architectures.

\begin{figure}
    \centering
    \begin{minipage}{0.3\linewidth}
    \begin{small}
    \begin{tabular}{l}
         \textbf{Category labels}\\
         \hline
         0-"Cable/DSL/ISP"\\
         1-"Content"\\
         2-"Enterprise"\\
         3-"NSP"\\
         4-"Not Disclosed"\\
         5-"Other"
    \end{tabular}
    \end{small}
    \end{minipage}
    \hspace{0.08\linewidth}
    \begin{minipage}{0.6\linewidth}
    \includegraphics[width=1\linewidth]{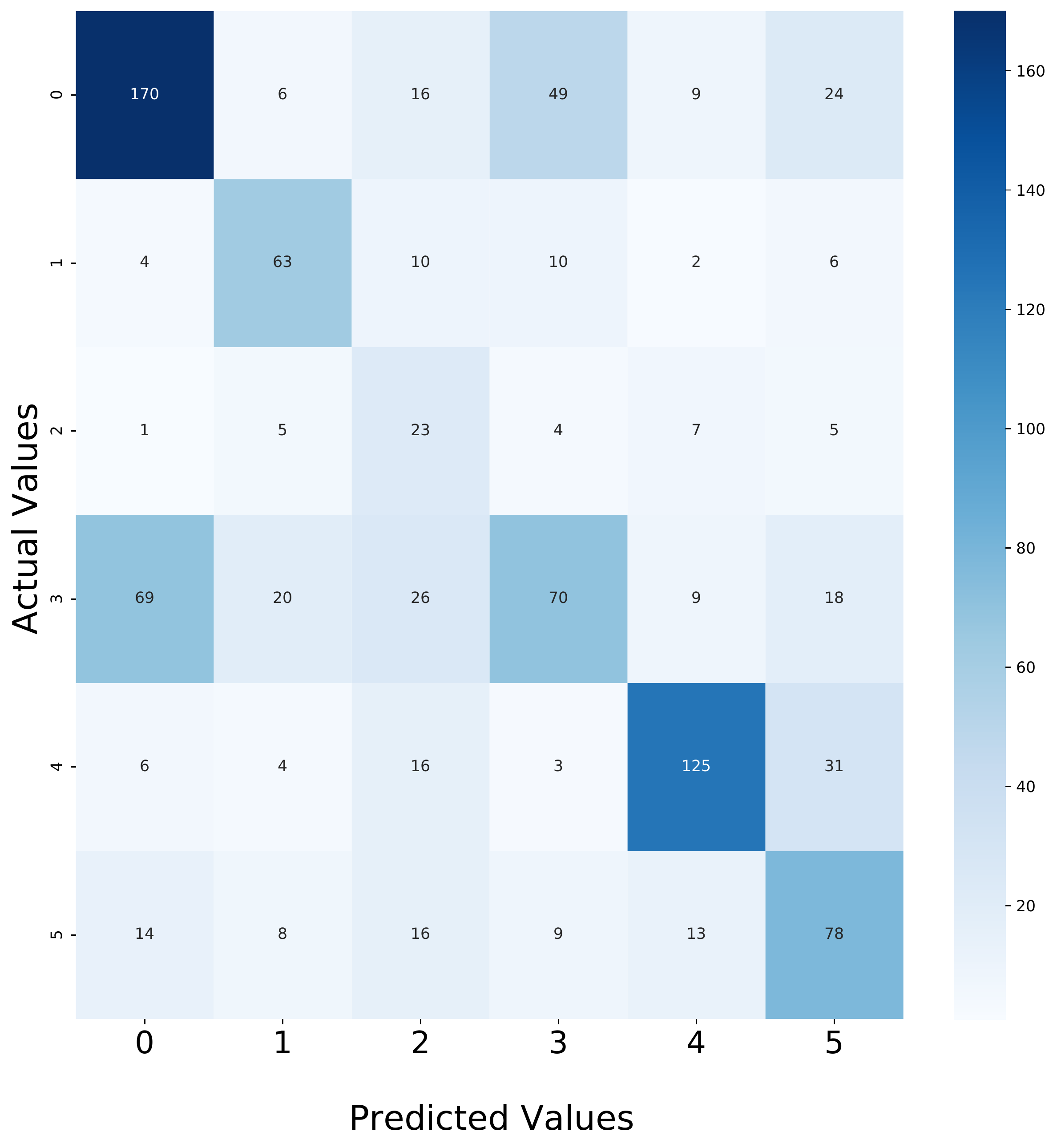}
    \end{minipage}
    \caption{Heatmap of predicted vs. actual values of the GraphSAGE model for the "Network type" attribute. %The numeric labels correspond to the categories: 0-"Cable/DSL/ISP", 1-"Content", 2-"Enterprise", 3-"NSP", 4-"Not Disclosed", 5-"other".
    }
    \label{fig:heatmap_pdb}
\end{figure}

As an example, Fig.~\ref{fig:heatmap_pdb} depicts the detailed predictions of the GraphSAGE model for the "Network type" attribute. In all categories (rows), the number of samples (cell values) that are predicted correctly (diagonal) are higher than in any other category (columns) in the same row. The most missclassifications are between ASes that are characterized "Cable/DSL/ISP" and "NSP (Network Service Providers)", which in practice can have several common characteristics.

Compared to the RF model, GNNs predict better only the "Scope" and "Network type" attributes. This indicates that not all node attributes may be strongly related to the underlying graph structure; in some cases the graph structure can help our predictions, whereas in other cases using a simpler (i.e., easier to train) model may be the best solution. 

Finally, it is clear from the node2vec/bgp2vec performance that using only graph information is not enough for the node classification tasks. This further highlights the need for future research on applying GNNs (i.e., both graph structure and node attributes) for Internet routing related tasks.

\begin{table*}
    \centering
    \caption{Results for the node classification tasks: average accuracy (ACC) and F1 score metrics over 10 runs per model.}
    \label{tab:results-node-classification}
    \begin{tabular}{c|cc|cc|cc|cc}
         & \multicolumn{2}{c|}{\textbf{Traffic ratio (PDB)}} & \multicolumn{2}{c|}{\textbf{Scope (PDB)}}& \multicolumn{2}{c|}{\textbf{Network type (PDB)}}& \multicolumn{2}{c}{\textbf{Peering policy (PDB)}}\\
         \textbf{Model} & ACC & F1 & ACC & F1& ACC & F1& ACC & F1\\
         \hline
         GraphSAGE  & 44.8\% & 35.9\% & 49.2\% & 47.2\% & 54.7\% & 53.1\% & 34.9\% & 30.6\% \\
         GCN        & 38.7\% & 30.5\% & 40.1\% & 37.1\% & 46.6\% & 44.6\% & 37.0\% & 31.1\% \\
         GAT        & 38.0\% & 31.3\% & 41.8\% & 38.4\% & 49.9\% & 47.2\% & 32.2\% & 28.8\% \\
         % GIN        & 29.0\% & 09.3\% & 10.1\% & 05.1\% & 29.0\% & 10.2\% & 60.0\% & 25.0\% \\
         \hline
         node2vec   & 20.9\% & 19.4\% & 16.3\% & 15.7\% & 19.0\% & 18.6\% & 31.1\% & 27.0\% \\
         bgp2vec    & 14.8\% & 13.4\% & 14.8\% & 13.0\% & 19.9\% & 19.1\% & 29.4\% & 26.8\% \\
         \hline
         Rnd. Forest& 51.1\% & 35.8\% & 36.6\% & 33.7\% & 49.8\% & 42.9\% & 54.6\% & 34.8\% \\
    \end{tabular}
\end{table*}

%\myitem{The role of class imbalance \& small classes.} In the dataset, for some categorical node features, there are classes with very few samples \pavlos{(we provide visualizations of the detailed distributions of all features -to facilitate the exploratory data analysis- in REPO-LINK)}. Neglecting the class imbalance leads to high accuracy, however, all models tend to predict only the main class(es). Of course, this is expected and there are several methods to overcome in ML. However, trying to balance the training dataset by selecting equal number of samples per class, leads to very few samples (because the size of the smaller classes is very small, e.g., less than a hundred samples), which was insufficient for training complex GNN models. That is one of the main reasons for GNN models underperforming baseline models that do not capture graph structure (such as, random forests). .

\section{Related Work}

In the last few years, there have been several attempts for combining Internet routing data with ML, with the majority of them focusing on BGP anomaly detection. Ding et al.~\cite{ding2016BGPanomalies} and Dai et al.~\cite{dai2019-ml-bgp-anomalies} try to detect BGP anomalies using traditional ML models, various features, and advanced feature selection methodologies, such as minimum redundancy maximum relevance~\cite{ding2016BGPanomalies} or Fisher linear analysis and Markov random fields~\cite{dai2019-ml-bgp-anomalies}

Typically, for Internet routing tasks, features can be extracted from BGP messages~\cite{fonseca2019bgp, hoarau2021graph-representation-bgp, testart2019profiling, Cho2019BGPHC, sermpezis2021estimating}, for example, volume, AS path features and BGP attributes~\cite{fonseca2019bgp, hoarau2021graph-representation-bgp}, network importance metrics~\cite{Cho2019BGPHC}, network observations~\cite{sermpezis2021estimating}, or graph-level metrics~\cite{hoarau2021graph-representation-bgp}. Sanchez et al.~\cite{sanchez2019comparing-ml-bgp} went a step further to consider more robust graph features, such as node centrality, clique theory, etc. They conclude in the fact that centrality metrics  are more likely to detect large-scale incidents. 

The first work that proposed the use of graph embeddings on ASN level is~\cite{shapira2022bgp2vec,shapira2020hijack-asn-embedding}, which uses BGP messages and the AS-paths in them to propose the bgp2vec model. These embeddings can be used to predict node and link properties~\cite{shapira2022bgp2vec} or classify BGP routes as standard or hijacked~\cite{shapira2020hijack-asn-embedding}.

Finally, some recent  benchmarking efforts for GNNs in the domain of networking (but not for Internet routing data) are (i) the IGNNITION~\cite{perich2021ignnition} framework for prototyping GNNs for communication networks, which contains tools to design, train and evaluate a GNN model, and (ii) the GNNet challenge~\cite{suarez2021graph} for designing GNN models for predicting network performance.

\section{Conclusion}
In this paper we compiled a benchmark dataset with Internet data, preprocessed in a way that is compatible to be used by GNN architectures. The benefits from the dataset are twofold: (i) it enables researchers to focus on designing GNN architectures rather than collecting and processing data, which can be a time consuming task (or even prohibitive for non Internet experts), and (ii) it can serve as a common dataset where different works can be compared on (following the example of~\cite{suarez2021graph}), which is a key requirement for progressing ML research for networking~\cite{casas2020two}. 

Using our dataset and pipeline, we performed various experiments to test the performance of different GNN models. Our initial results, not only can be used as a baseline in future work, but also provide useful insights. We showed that capturing both node attributes and graph structure can be beneficial for link prediction and node classification tasks, however, the benefits of each factor may vary depending on the problem. Given the lack of previous works with GNNs on Internet data, and thus the lack of reported insights, we believe these findings can be a useful starting point for future research.

% \pavlos{There are several directions that can be considered in the future, including optimization of GNN architectures, design of GNN architectures specifically for the, investigation of the role of how the AS-graph is built (e.g., it is known that the AS paths from which we generate the graph are collected by specific vantage points with known biases: increasing/decreasing the bias could affect our results; or similarly, increasing/decreasing the completeness of the graph, e.g., with more vantage points), if edge attributes could further improve performance (i.e., peering relationships types in the CAIDA AS-relationships dataset). Also our results about the different link accuracies, could be generalized about the fairness of models for node classificiation as well, etc.}


\begin{acks}
% \section*{Acknowledgements}
This research is co-financed by Greece and European Union through the Operational Program Competitiveness, Entrepreneurship and Innovation under the call RESEARCH-CREATE-INNOVATE (projects T2EDK-04937 and T2EDK-03898), the European High-Performance Computing Joint Undertaking (GA No. 951732), and RIPE NCC (AI4NetMon project).
\end{acks}
\bibliographystyle{ACM-Reference-Format}
\bibliography{references}

% \appendix

\end{document}